\documentclass[9pt,twocolumn,twoside]{opticajnl}
\journal{opticajournal} 

\setboolean{shortarticle}{true}

\usepackage[inkscapelatex=false]{svg}
\usepackage{lineno}
\usepackage{graphicx}
\usepackage{subcaption}
\usepackage{float}
\usepackage{placeins}

\title{Multi-core anti-resonant hollow core optical fibre}

\author[1,*]{Robbie Mears}
\author[]{Kerrianne Harrington}
\author[]{William J. Wadsworth}
\author[]{James M. Stone}
\author[]{Tim A. Birks}

\affil{Centre for Photonics and Photonic Materials, Department of Physics, University of Bath, Bath, BA2 7AY, UK}

\affil[*]{rm2033@bath.ac.uk}

\begin{abstract}
We report the fabrication and characterisation of a multi-core anti-resonant hollow core fibre with low inter-core coupling. The optical losses were 0.03 and 0.08 dB/m at 620 and 1000 nm respectively, while the novel structure provides new insights into hollow core fibre design and fabrication.
\end{abstract}

\setboolean{displaycopyright}{false} 

\begin{document}

\maketitle

\section{Introduction}
Multi-core fibres are widely used across a range of applications, from optical fibre sensing to spatial division multiplexing \cite{Puttnam21}. With a high degree of flexibility in their design and fabrication, multi-core fibres can easily be designed to be single moded or few moded as required. The use of existing single core infrastructure has enabled their rapid development and deployment in research and industry, as demonstrated by the adoption of multi-core fibres in the telecommunications field \cite{Winzer18}.

Another rapidly developing field is that of hollow core fibres \cite{cregan1999,sakr2020,Chen24}, which guide light in air or vacuum. This can be achieved through various types of microstructure which confine the light to a central hollow core, with a greatly reduced overlap of the light with the fibre material. With material contributions to loss, dispersion and damage thresholds heavily suppressed, hollow core fibres allow shorter and more intense pulses of light to be employed across a wider spectral range. With measured losses in state of the art hollow core fibres now beating silica fibres at telecommunications wavelengths \cite{Chen24}, hollow core fibres are an emerging technology in a range of fields previously hindered by the limitations of solid fibres \cite{Ding20}.

One iteration of hollow core fibres, sometimes called tubular anti-resonant fibres, achieves confinement through a number of thin glass resonators around a central air core as illustrated in Figure \ref{fig:thick_fibres} (d). While a fundamentally leaky confinement mechanism, as light is lost on each reflection from the core boundary, mechanisms such as azimuthal confinement \cite{Murphy23} and anti-resonance \cite{bird2017} reduce the loss far below that possible in a simple capillary waveguide. Additional control of the air regions of these resonators allow the enhanced leakage of higher order modes\cite{Uebel16}, to create effectively single mode fibres despite being strictly multi-mode. Using these principles, anti-resonant fibres can be fabricated to guide at wavelengths from the vacuum-ultraviolet \cite{Winter19} to the mid-infrared \cite{YuIR}, far surpassing the range possible in solid silica fibres.

Despite the current interest in solid multi-core fibres and hollow core fibres, there have only been a few recent examples of multi-core hollow core fibres. In hollow core photonic-bandgap fibres, additional 'shunt' cores have been long employed \cite{Fini13} to introduce additional loss for higher order modes. However, to our knowledge there has only been one example \cite{mangan2010} of a multi-core hollow-core bandgap fibre where all cores are intended to guide with low loss. This 7 core example utilised separate microstructures for each core, with no coupling reported. Within Kagome type structures, there have been a few reports of multiple cores in a common  microstructure, with coupling reported to depend on their separation and polarisation state\cite{pmma,wheeler2016,Rodrigo24}. For the newer and more effective tubular design, 6 resonator fibres with a single resonator shared between each core, forming two \cite{twocore17} or three air cores \cite{threecore18} have been reported. This design utilised the coupling between cores to operate as a fibre coupler and interferometer \cite{twocore17,interfere2019}, but the propagation loss was not measured. 

In this work we demonstrate the uncoupled variant, drawing a three core fibre where coupling is reduced greatly by using separate microstructures for each core. We describe the fabrication challenges and considerations associated with drawing a multi-core fibre, particularly the displacement of the air core from the centre of the fibre. The characteristics of a three core fibre are probed, and we measure the propagation loss of the individual cores along with core to core coupling. We demonstrate that low loss uncoupled guidance in a multi-core hollow core fibres is possible. Through the unusual structure we observe a new coupling phenomenon, not previously reported to our knowledge, that may be applicable to anti-resonant fibres more generally.
\section{Fabrication}
\begin{figure}[h]
\centering
\fbox{\includegraphics[width=0.9\linewidth]{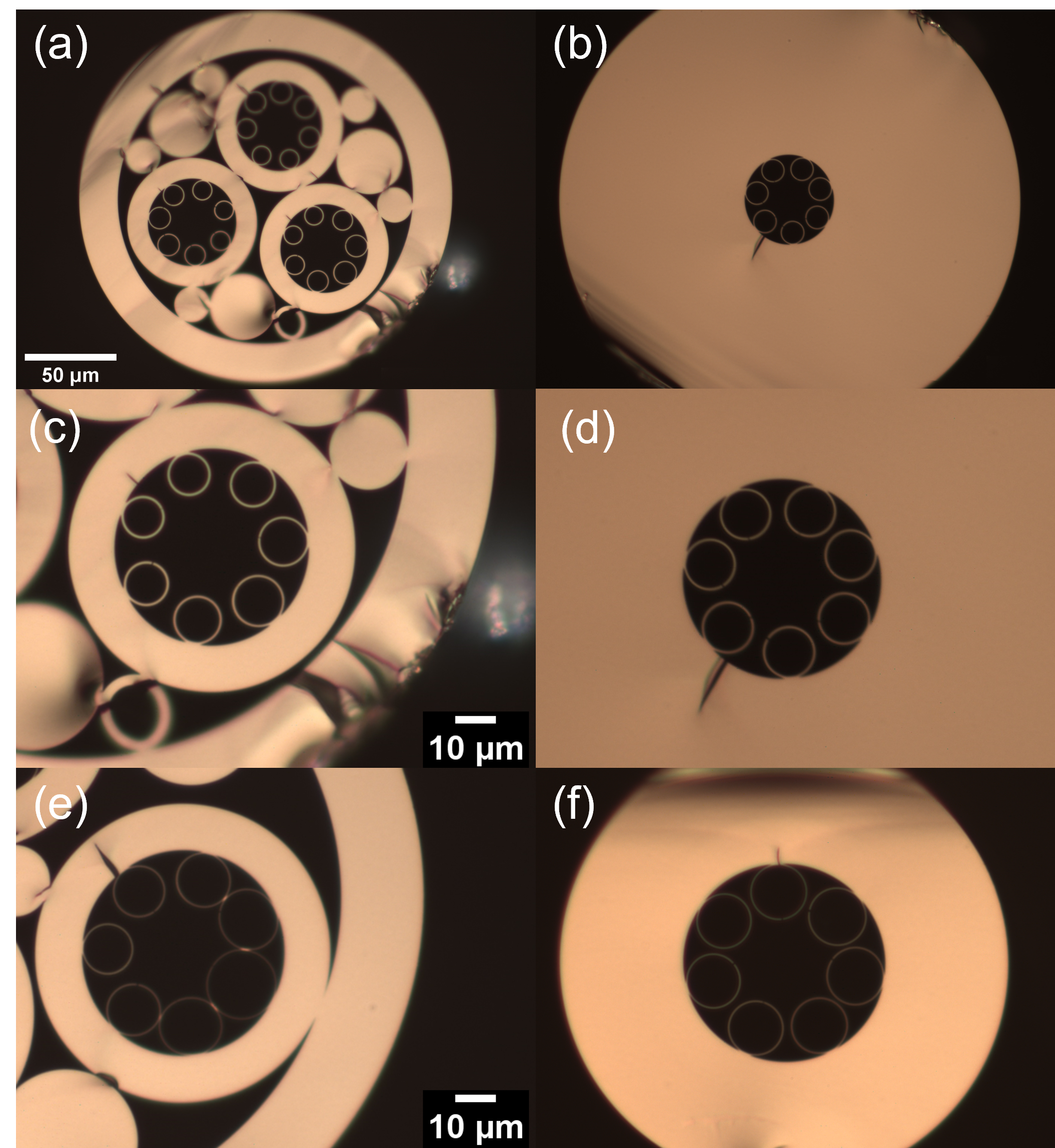}}
\centering
\caption{A series of optical micrographs comparing the structure of multi-core fibres to single core fibres. (a) The reported multi-core hollow core fibre, with a marker capillary at the bottom-right. (b) A single core hollow core fibre, using the same batch of canes as the reported fibre. (c) A magnified view of the bottom right core of (a), with the marker capillary included for reference. (d) The single core of (b) expanded for comparison. (e) A single core from another multi core fibre with thinner resonators than (a). (f) A single core fibre drawn using the same canes as (e). Horizontal pairs of images are to the same scale.}
\label{fig:thick_fibres}
\end{figure}

The fabrication process began with the typical stack and draw method \cite{Murphy22}, only differing towards the fibre drawing stage. Resonators of 2745 ± 3 µm outer diameter were drawn from a tube of 17/25 mm (inner and outer diameter) ratio. These were stacked inside another 17/25 mm tube and drawn to canes of 2850 ± 5 µm outer diameter. Three canes were then finally stacked inside a tube of 6.4/8 mm with structural supports before being drawn to fibre. The fibre was drawn to a 200 µm outer diameter, representing a draw-down ratio of 40 and a potential yield of 1.6 km for a 1 m preform. In this draw the parameters used were a set furnace temperature of 1920 °C, a preform feed rate of 35 mm/min, draw speed of \(\sim\) 53 m/min and a drawing tension of \(\sim\) 360 g. To achieve the desired resonator sizes, we apply a differential pressure between the core and resonator regions, which serves to inflate the resonators during the drawing process. A parameter sweep in pressure yielded 150 m of the fibre in Figure \ref{fig:thick_fibres} (a) with a uniform differential pressure of 34 kPa applied to all resonators.

In this fibre, and other similar multi-core fibres fabricated, we observed a gradient of resonator inflation within the drawn fibre, with those nearer the outside of the preform experiencing greater inflation than those at the centre, Fig \ref{fig:thick_fibres}. This persistent asymmetry suggests a higher temperature towards the outside of the preform, leading to a difference in viscosity across the preform cross-section during drawing and hence greater inflation of the pressurised resonators nearer the outside of the preform. To confirm that any resonator variation in our drawn canes or errors of our drawing tower was not responsible for this effect, we drew a single cane from the same stack in a jacket of 3/10 mm to 250 µm, Fig \ref{fig:thick_fibres} (b), representing the same draw-down ratio. The fibre was drawn using a furnace temperature of 1900 °C, a feed rate of 25 mm/min, a draw speed of 40 m/min, drawing tension of 685 g and a differential resonator pressure of 26 kPa. While the drawing parameters differ from those used in the multi-core fibre fabrication, the similar resonator inflation and pressure would suggest similar levels of microstructure control. Given the lack of comparable asymmetry in the single core fibre compared to the multi-core variation we conclude that the asymmetry in our multi-core fibres is induced by the breaking of rotational symmetry that is usually present in a single core fibre. A similar gradient in inflation was also observed in the multi-core hollow core photonic bandgap fibre \cite{mangan2010}, providing further evidence of a systematic effect.

This effect poses a significant and fundamental limitation on the structural control within these multi-core fibres, because rapidly inflating outer resonators exhibit mid-draw contact \cite{Jasion19} before more central resonators. While asymmetry of inflation is often observed in single core hollow core fibres due to a variety of fabrication errors, these can be minimised to create highly uniform structures. In these multi-core structures the asymmetry is linked to the resonator position and cannot be similarly controlled. As a result, the final structure has larger inter-resonator gaps and a wider range of wall thicknesses, both known to increase fibre losses. While the asymmetry is not severe in this case, other multi-core fibres fabricated with thinner walls as shown in Figure \ref{fig:thick_fibres} (e) exhibited greater asymmetry which severely limited their optical performance. This is again in contrast to the single core fibre drawn from the same canes, shown in Figure \ref{fig:thick_fibres} (f), which show minimal resonator asymmetry.
\section{Characterisation}
\begin{figure}[ht]
\centering
\includegraphics[width=1\linewidth]{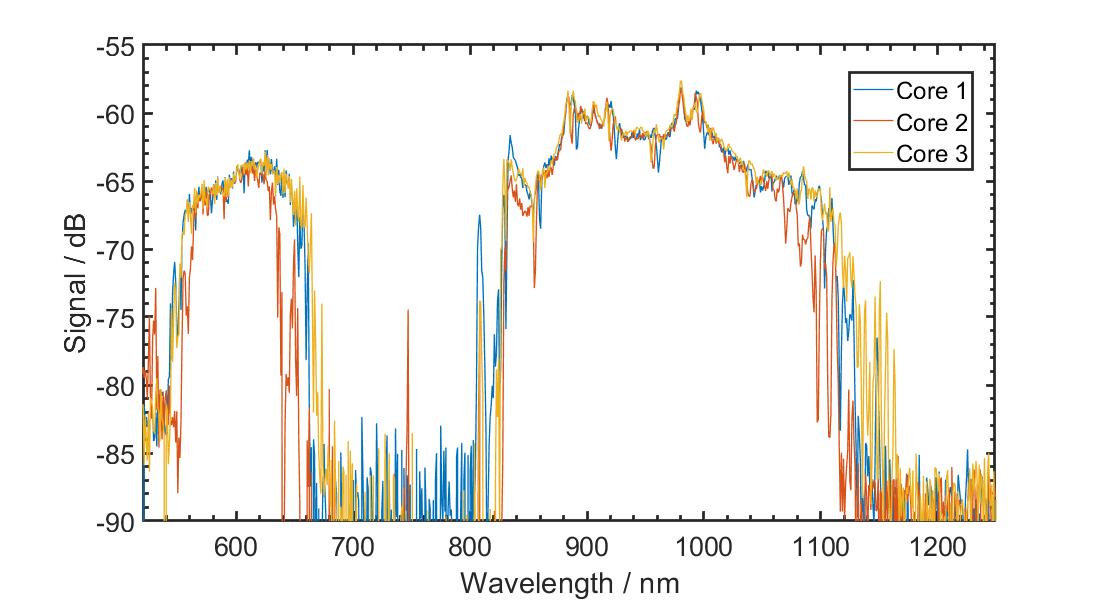}
\caption{Transmission spectra on a logarithmic scale for the three different cores of the multi-core fibre. }
\label{fig:transmission}
\end{figure}

The experimental setup used for optical characterisation of the fibre consisted of an Energetiq EQ-99X laser-driven white light source and ANDO Optical Spectrum Analyser, with single mode fibre (SMF-28) used at the input to selectively couple to one core at time within the fibre under test. While SMF-28 is few moded in the wavelengths guided by the multi-core fibre, the 7 ring design is effectively single moded after a short distance, with a single mode pattern observed at the output after only a few metres. To compare the uniformity between cores, we first tested the transmission of each core over a short length \(\sim\) 3 m, using a saturated near field image to distinguish each core relative to the marker capillary. In Figure \ref{fig:transmission} we show the transmission through the three separate cores.

\begin{figure}[h]
\centering
\fbox{\includegraphics[width=1\linewidth]{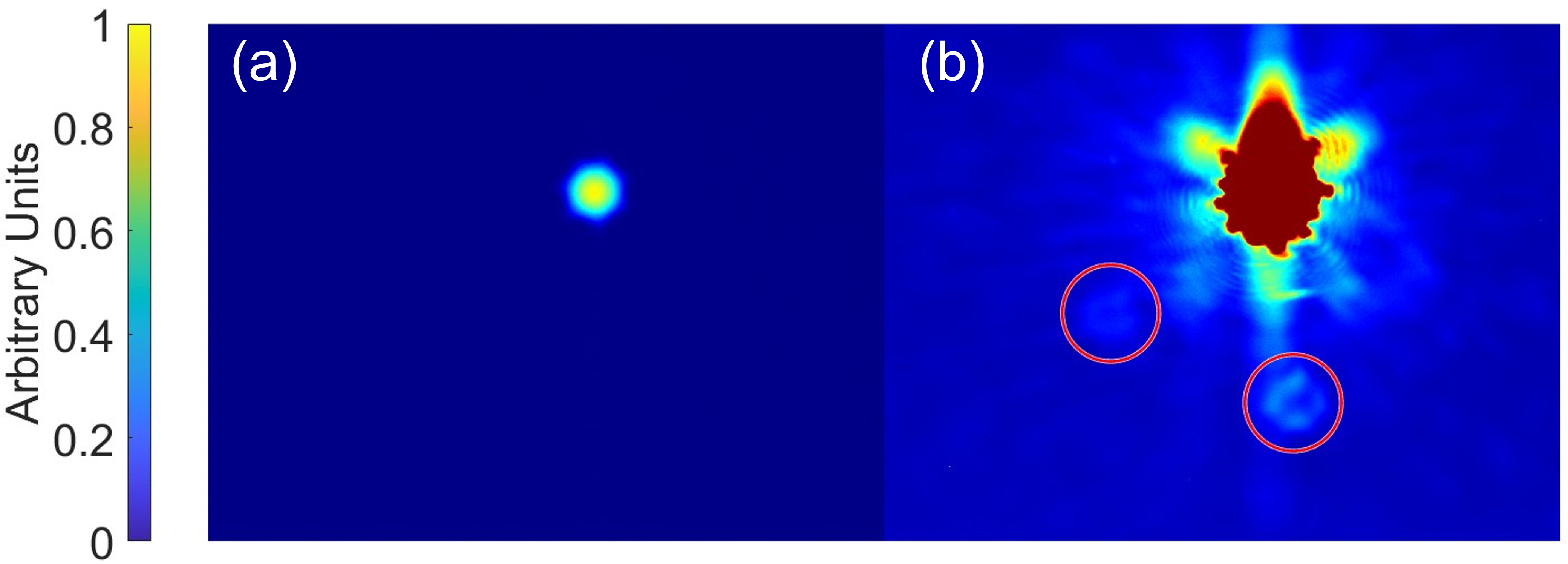}}
\centering
\caption{(a) Unsaturated near field image of a single core at the output of 47.7 m of the multi-core fibre. (b) Near field image with \(\sim\) 20,000 x greater exposure than (a). The two cores not initially excited are circled in red, showing a small amount of coupled light. 
}
\label{fig:nearfield1}
\end{figure}

\FloatBarrier
In the transmission spectrum of a tubular anti-resonant fibre we expect a series of high loss resonances separated by broad anti-resonant bands with low loss. The variation in resonator size is expected to broaden these resonances as the wider range of resonator thicknesses will have different resonant conditions. This is shown in Figure \ref{fig:transmission}, where the broad resonant regions are consistent with the \textit{m} = 1 resonance at \(\sim\) 1500 nm and the \textit{m} = 2 resonance at \(\sim\) 750 nm where \textit{m} is the order of the resonance of the wall. From the \textit{m} = 2 resonance centred on \(\sim\) 750 nm, an estimated wall thickness of \(\sim\) 700 nm is calculated using the ARROW model \cite{litchinitser2002}. The lack of fine structure within the transmission bands over this short distance, present in other multi-core examples \cite{Jia21} due to inter-core beating, would suggests a lack of strong coupling between the cores. To further investigate the inter-core coupling we imaged the near field output after 47.7 m of fibre using a single core excitation, and saturated the silicon camera to show that only a small amount of light is present outside the initially excited core, as shown in Figure \ref{fig:nearfield1} (b).

We additionally tested the effect of bending on coupling between the cores, we observed similar coupling to Figure \ref{fig:nearfield1} (b) for bends down to  \(\sim\) 2 cm radius. In all cases, the coupled light within other cores appear as a LP\(_{11}\) like mode. This can be rationalised by considering that bend loss in anti-resonant fibres is expected to take place through coupling to resonator air modes \cite{carter2017} which are strongly coupled to the solid cladding. In this multi-core structure, the cladding hosts other resonators and cores with similar propagation constants, allowing a certain degree of coupling. However, these LP\(_{11}\) modes are inherently high loss in a six or seven ring fibre through strong coupling to the fundamental resonator air mode \cite{Uebel16}. When combined with the large physical separation of cores, this results in a low degree of inter-core coupling. Based on the near field image in Figure \ref{fig:nearfield1} the power in the other cores is \(\sim\) 40 dB below the excited core after propagation through 47.7 m of the fibre.

\begin{figure}[h]
\centering
\includegraphics[width=1\linewidth]{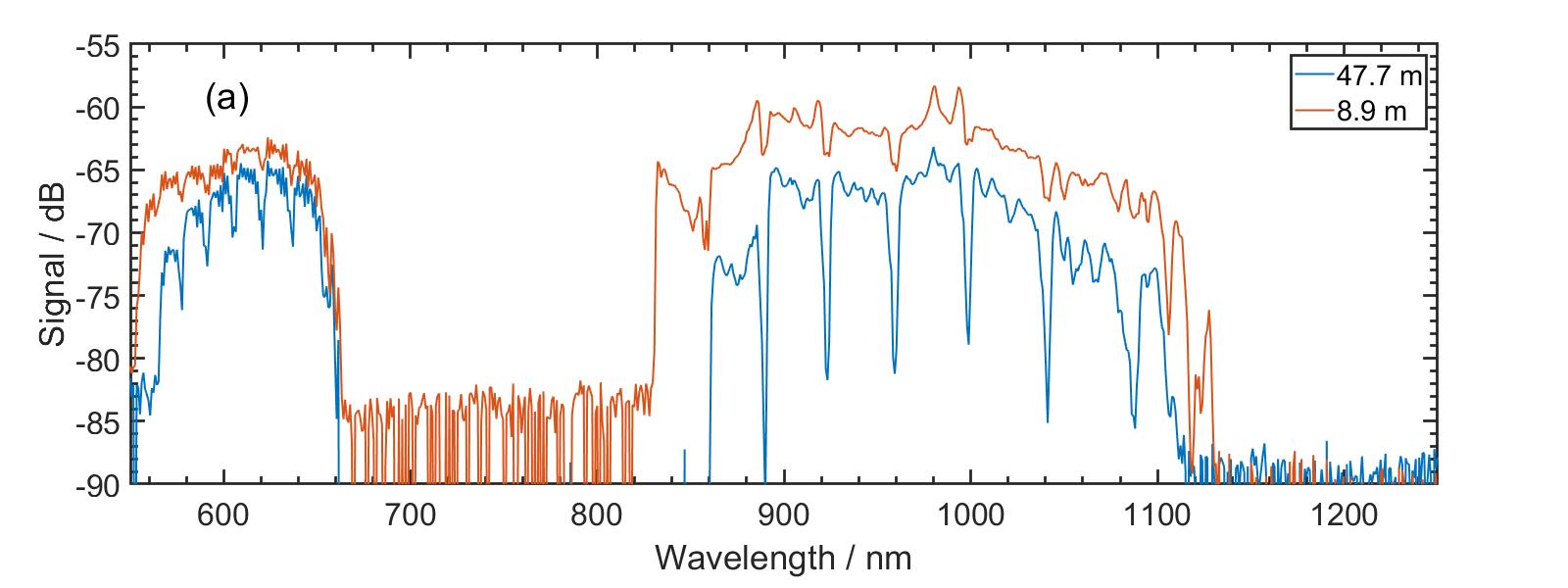}
\centering
\includegraphics[width=1\linewidth]{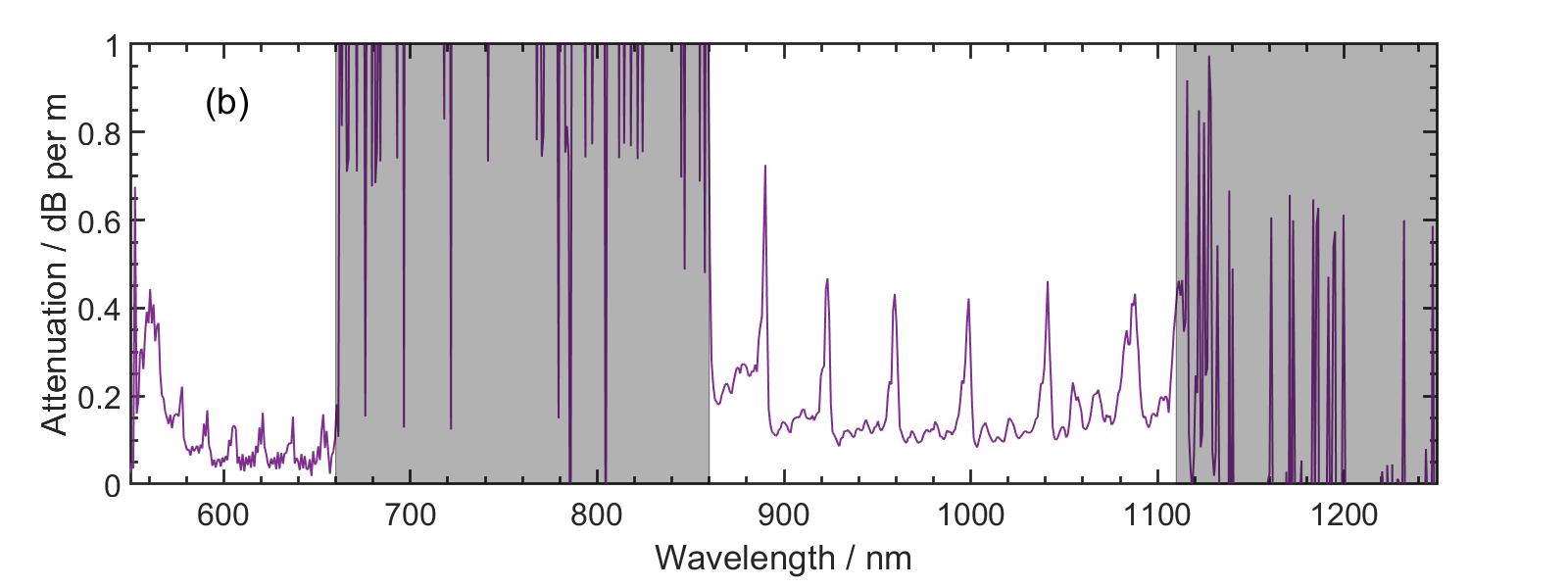}
\centering
\includegraphics[width=1\linewidth]{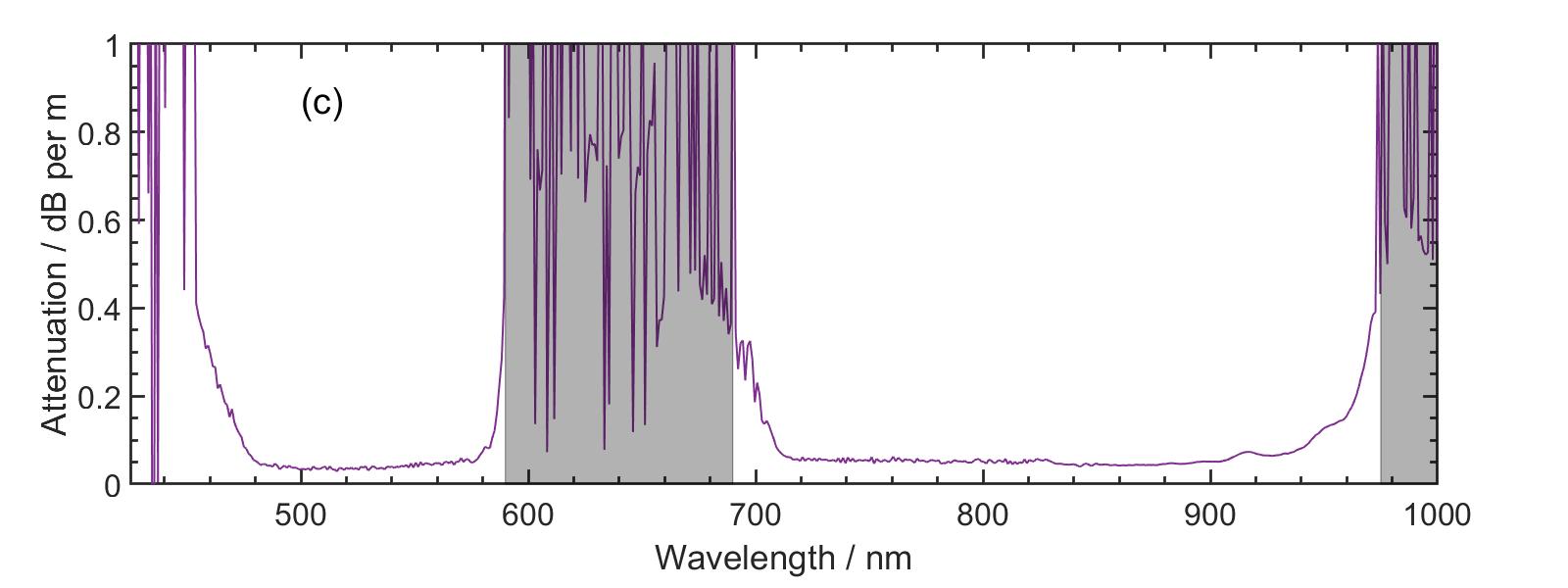}
\caption{(a) Transmission spectra versus wavelength on a logarithmic scale for fibre lengths of 47.7 m and 8.6 m for a single excited core. (b) Attenuation calculated from the above transmission spectra. Shaded regions indicate no signal above noise (c) Attenuation of the single core fibre from Figure \ref{fig:thick_fibres} (b) for a cutback from 49.9 m to 8.9 m.}
\label{fig:cutback}
\end{figure}

The 47.7 m length of fibre was then laid in loose loops of \(\sim\) 1 m diameter with a single core excitation as before. The excitation was moved between the three cores, with similar transmission spectra, suggesting that the losses through the multiple cores are comparable. Then for a single core we performed multiple transmission scans with the fibre output re-cleaved on each scan, followed by a cutback to 8.9 m where the same process was repeated. In Figure \ref{fig:cutback} (a) we show the transmission through both lengths, with Figure \ref{fig:cutback} (b) showing the calculated attenuation spectrum across that spectral range. Over a short length additional transmission bands in the visible and ultraviolet were observed, but over the longer length these did not persist.

The minimum attenuations determined fro the cutback measurements were 0.03 and 0.08 dB/m at 620 and 1000 nm respectively, with sharp increases in loss at regular intervals within both transmission bands. These loss features are clearly visible on the transmission plots in Figure \ref{fig:cutback} (a) with their position and amplitude invariant with length or bending. This behaviour would suggest a coupling or loss mechanism in the transverse direction of the fibre rather than multimode beating along the fibre. Given the unusual structure of the cladding, we propose that these features are due to a lossy resonant coupling with the cane jacket, which surrounds each core. Using the ARROW model \cite{litchinitser2002} for a silica layer in air it can be readily shown that the separation of resonances in wavevector \(\Delta k\) is given by:

\begin{equation}
\Delta k = \frac{\pi}{t(n^2-1)^{\frac{1}{2}}}  \label{eqn1}
\end{equation}
where \textit{t} is the resonant layer thickness and \textit{n} is the refractive index of silica. The separation of the loss peaks were analysed using Equation \ref{eqn1} and a thickness of 10.4 µm was calculated. This agrees well with the measured jacket wall thickness of \(\sim\) 11 µm in the drawn fibre. This effect was not seen in the typical single core example, as shown in Figure \ref{fig:cutback} (c).

Due to the previously discussed gradient of inflation, the range of resonator thicknesses serves to broaden the resonant regions as expected, reducing the low loss bandwidth. This is highlighted when compared to the flat attenuation of the single core example, shown in Figure \ref{fig:cutback} (c), where a similar attenuation is achieved across a much wider bandwidth with narrower resonance regions. The cladding coupling that occurs in the multi-core fibre raises the attenuation by \(\sim\) 0.3 - 0.6 dB/m at the resonant wavelengths of the jacket wall, compared to the general background trend of attenuation. While the increased loss at the resonances is undesirable, an anti-resonant suppression of loss is likely to exist between the resonances, reducing the confinement loss compared to an all solid jacket. However due to the structural imperfections and differences between the two fibres we cannot confirm if this is the case.
\section{Discussion}
Through the fabrication of these multi-core fibres it appears that a temperature gradient exists within the fibre preform during the drawing process. While this limited the final quality of the drawn multi-core fibre, it also provides useful insight in the fabrication process. To our knowledge, a gradient in the transverse direction has not previously been reported, with most works focusing on the variation in the longitudinal direction \cite{Jasion19,Kostecki14}. In fact, in these works a uniform radial temperature distribution is assumed within the preform. Despite this, these models are accurate as the drawing process can be modelled using only the measured drawing tension \cite{Chen13}, independent of the actual neck-down profile or viscosity. The presence and effect of such a temperature gradient in standard single core structures should be further explored, particularly when considering that a higher stress on the central microstructure is advantageous \cite{Jasion19} compared to a uniformly distributed stress.

An unexpected outcome of this work has been the cladding resonance effect, made possible by the unusual cladding structure. The jacket tube supporting the resonators appears to be responsible for an additional family of loss peaks, enabled by the few points of contact between the tube and the rest of the fibre. This may provide an anti-resonant effect in various transmission bands, reducing the fibre confinement losses compared to an all-solid jacket. To maximise the available low loss bandwidth, the cane jacket thickness should be reduced to a similar thickness to the resonators.

\section{Conclusion}
We have demonstrated a low loss three core hollow core fibre, with guidance in the visible and near infra-red. A low level of inter-core coupling was shown, indicating that largely uncoupled multi-core hollow core fibres are possible by using separate microstructures for each core. The breaking of rotational symmetry in the reported design has demonstrated a previously overlooked temperature gradient during the fibre drawing process, which may be important for future fibre fabrication, particularly those with complex microstructures. Finally, a new cladding anti-resonant effect is shown, which introduces another tool for hollow core fibre design and fabrication.
\section{Backmatter}
\begin{backmatter}
\bmsection{Funding}This work was funded by the EPSRC under grant EP/T020903/1.

\bmsection{Disclosures} The authors declare no conflicts of interest.

\bmsection{Data Availability Statement} The relevant data is available from \cite{data}.

\end{backmatter}
\bibliography{sample}
\bibliographyfullrefs{sample}

\end{document}